\documentclass[10pt,twoside]{IEEEtran}

\usepackage{etoolbox}
\newtoggle{doublecolumn}
\newtoggle{calculateFigures}

\toggletrue{doublecolumn}
\togglefalse{calculateFigures}

\usepackage{etex}

\usepackage[english]{babel}
\usepackage{lipsum}
\usepackage{amsbsy,amsmath,amsfonts,amssymb,amsthm}
\usepackage{mathtools}
\usepackage{textcomp} 
\usepackage{relsize}

\usepackage{subdepth} 

\usepackage{hyperref}

\usepackage{bm,cite}
\usepackage{cases}
\usepackage{times,url,verbatim}

\usepackage[noend]{algpseudocode}
\usepackage{booktabs}
\usepackage{tabularx,array,dcolumn,multirow}

\usepackage{enumitem} 

\usepackage{algorithm}

\usepackage{yfonts}

\usepackage{xr} 

\usepackage{soul} 

\usepackage{blkarray,bigdelim} 

\allowdisplaybreaks[4]

\usepackage{centernot} 

\newcommand{\subparagraph}{}

\iftoggle{doublecolumn}{
    \newtheorem{thm}{Theorem}
    \newtheorem{fact}{Fact}
    \newtheorem{lemma}{Lemma}
    \newtheorem{definition}{Definition}
    \newtheorem{conj}{Conjecture}
    \newtheorem{propos}{Proposition}
    \newtheorem{corol}{Corollary}
    \newtheorem{ass}{Assumption}
    \newtheorem{example}{Example}
    \newtheorem{remark}{Remark}
    \newtheorem{note}{Note}
    \newtheorem{obs}{Observation}
}{

    \newtheoremstyle{exampstyle}
      {0} 
      {0} 
      {\itshape} 
      {} 
      {\bfseries} 
      {.} 
      {.5em} 
      {} 

    \theoremstyle{exampstyle} 
    \theoremstyle{exampstyle} 
    \theoremstyle{exampstyle} 
    \theoremstyle{exampstyle} 
    \theoremstyle{exampstyle} 
    \theoremstyle{exampstyle} 
    \theoremstyle{exampstyle} 
    \theoremstyle{exampstyle} 
    \theoremstyle{exampstyle} 
    \theoremstyle{exampstyle} 
    \theoremstyle{exampstyle} 
    \theoremstyle{exampstyle} 
}

\newcommand{\argmax}[1]{\underset{#1}{\operatorname{arg}\,\operatorname{max}}\;}

\makeatletter
\newcommand{\pushright}[1]{\ifmeasuring@#1\else\omit\hfill$\displaystyle#1$\fi\ignorespaces}
\newcommand{\pushleft}[1]{\ifmeasuring@#1\else\omit$\displaystyle#1$\hfill\fi\ignorespaces}

\begingroup
\catcode`\#=11

\endgroup

\usepackage{graphicx,xcolor,float,dblfloatfix}
\usepackage{psfrag}

\usepackage{caption}
\usepackage{subcaption}

\usepackage[compact]{titlesec} 

\newcommand{\subalign}[1]{%
  \vcenter{%
    \Let@ \restore@math@cr \default@tag
    \baselineskip\fontdimen10 \scriptfont\tw@
    \advance\baselineskip\fontdimen12 \scriptfont\tw@
    \lineskip\thr@@\fontdimen8 \scriptfont\thr@@
    \lineskiplimit\lineskip
    \ialign{\hfil$\m@th\textstyle##$&$\m@th\textstyle{}##$\crcr
      #1\crcr
    }%
  }
}

\graphicspath{%
{./Figures/}
}

\usepackage{color}

\usepackage{caption}
\usepackage{subcaption}
\usepackage[compact]{titlesec}

\begin{document}

\title{Decentralized Transmission Policies\\ for Energy Harvesting Devices}

\author{Alessandro~Biason\IEEEauthorrefmark{1}, Subhrakanti~Dey\IEEEauthorrefmark{2} and Michele~Zorzi\IEEEauthorrefmark{1}\\
\IEEEauthorblockA{\small \IEEEauthorrefmark{1} Department of Information Engineering, University of Padova - via Gradenigo
6b, 35131 Padova, Italy}\\
\IEEEauthorblockA{\small \IEEEauthorrefmark{2} Department of Engineering Science, Uppsala University, Uppsala, Sweden}\\
\IEEEauthorblockA{\small email: biasonal@dei.unipd.it, Subhrakanti.Dey@signal.uu.se, zorzi@dei.unipd.it
\vspace{-1cm}}
\thanks{An extended version of this paper can be found in~\cite{Biason2017SubhraJ}.}
}

\maketitle
\pagestyle{empty}
\thispagestyle{empty}

\fontdimen2\font=3.1pt

\begin{abstract}
The problem of finding decentralized transmission policies in a wireless communication network with energy harvesting constraints is formulated and solved using the decentralized Markov decision process framework. The proposed policy defines the transmission strategies of all devices so as to correctly balance the collision probabilities with the energy constraints. After an initial coordination phase, in which the network parameters are initialized for all devices, every node proceeds in a fully decentralized fashion. We numerically show that, unlike in the case without energy constraints where a fully orthogonal scheme can be shown to be optimal, in the presence of energy harvesting this is no longer the best choice, and the optimal strategy lies between an orthogonal and a completely symmetric system.
\end{abstract}

\vspace{-0.2cm}
\section{Introduction}

Energy Harvesting (EH) has been established as one of the most prominent solutions for prolonging the lifetime and enhancing the performance of Wireless Sensor Networks (WSNs). Although this topic has been widely investigated in the literature so far, finding  proper energy management schemes is still an open issue in many cases of interest. In particular, using \emph{decentralized} policies, in which every node in the network acts autonomously and independently of the others is a major problem of practical interest in WSNs where a central controller may not be used all the time. Many decentralized communication schemes (e.g., Aloha-like) can be found in the literature; however, most of them were designed without a principle of \emph{optimality}, i.e., without explicitly trying to maximize the network performance. Instead, in this work we characterize the optimal decentralized policy in a WSN with EH constraints and describe the related computational issues. Although this approach intrinsically leads to a more complex protocol definition, it also characterizes the maximum performance a network can achieve, and may serve as a baseline for defining quasi-optimal low-complexity protocols.

Energy related problems in WSNs have been addressed by several previous works (see~\cite{Ulukus2015} and the references therein). Many analytical studies aimed at maximizing the performance of the network in terms of throughput~\cite{Michelusi2013,Biason2014,Tutuncuoglu2012a,Ozel2012b}, delay~\cite{Sharma2010}, quality of service~\cite{Pielli2016}, or other metrics. However, differently from this paper, most of the protocols proposed in the literature consider centralized policies, in which a controller coordinates all nodes and knows the global state of the system over time. \cite{Michelusi2015} analyzed decentralized policies with a particular focus on symmetric systems, and proposed a game theoretic approach for solving the problem. Instead, in this paper we use a different framework based on decentralized Markov decision processes, which can also handle asymmetric scenarios.

Recently, Dibangoye \emph{et al.}~\cite{Dibangoye2012,Dibangoye2013,Dibangoye2014} derived several important results in decentralized control theory. In this paper, we apply some of their results to an energy harvesting scenario, and, specifically, we model the system using a Decentralized-Markov Decision Process (Dec-MDP), which is a particular case of Decentralized-Partially Observable Markov Decision Process (Dec-POMDP). In \cite{Dibangoye2014}, a detailed study of the Dec-POMDPs was presented and different approaches to solve them were proposed. The notion of \emph{occupancy state} was introduced as a fundamental building block for Dec-POMDPs, and it was shown that, differently from classic statistical descriptions (e.g., belief states), it represents a sufficient statistic for control purposes. Using the occupancy state, we can convert the Dec-POMDP to an equivalent MDP with a continuous state space, named \emph{occupancy-MDP}. Then, standard techniques to solve POMDPs and MDPs can be applied; for example, an approach to solve a continuous state space MDP is to define a grid of points (see Lovejoy's grid approximation~\cite{Lovejoy1991}) and solve the MDP only in a subset of states. Although several papers introduced more advanced techniques to refine the grid~\cite{Zhou2001}, this approach may be inefficient and difficult to apply. Instead, in this paper we use a different scheme, namely the Learning Real Time A$^*$ (LRTA$^*$) algorithm~\cite{Korf1990}, which has the key advantage of exploring only the states which are actually visited by the process, without the difficulty of defining a grid of points.

Converting the Dec-POMDP to an occupancy-MDP produces a simpler formulation of the problem, which however does not reduce its complexity. Indeed, for every occupancy state, it is still required to perform the \emph{exhaustive backup} operation, i.e., to compute a decentralized control policy. This is the most critical operation in decentralized optimization, since it involves solving a non-convex problem with many variables. The problem can be simplified by imposing a predefined structure to the policy~\cite{Hoang2014}, so that only few parameters need to be optimized. While this may lead to suboptimal solutions, it greatly simplifies the numerical evaluation and, if correctly designed, produces close to optimal results.

The paper is organized as follows. Section~\ref{sec:system_model} presents the system model and the Dec-MDP formulation. The decentralized optimization problem is described in Section~\ref{sec:opt_problem} and solved in Section~\ref{sec:opt_solution}. The numerical results are shown in Section~\ref{sec:num_eval}. Finally, Section~\ref{sec:conclusions} concludes the paper.

\textbf{\emph{Notation.}} Throughout this paper, superscripts indicate the node indices, whereas subscripts are used for time indices. Boldface letters indicate \emph{global} quantities (i.e., vectors referred to all users).

\begin{figure}[!t]
  \centering
  \includegraphics[trim = 0mm 0mm 0mm 0mm, width=.95\columnwidth]{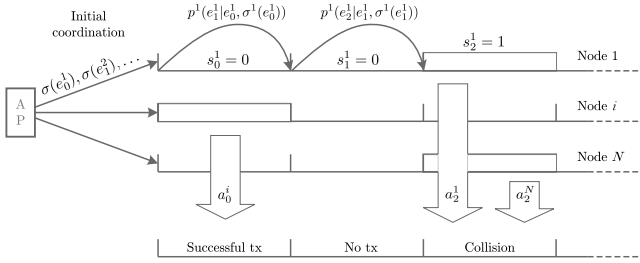}
  \caption{Time evolution of the system. After an initial coordination phase, every user acts independently of the others.}
  \label{fig:model}
\end{figure}

\section{System Model}\label{sec:system_model}

The network is composed of one Access Point (AP) and $N$ harvesting nodes (see \figurename~\ref{fig:model} for a graphical interpretation). 
We focus on a large time horizon, and time slot $k = 0,1,\ldots$ corresponds to the time interval $[kT,(k+1)T)$. During a slot, every node independently decides whether to access the uplink channel and transmit a message to AP, or to remain idle. We adopt an on/off collision model in which overlapping packet transmissions are always unrecoverable. 

In slot $k$, node $i$ harvests energy from the environment according to a pdf $B_k^i$ (e.g., similarly to~\cite{Michelusi2012}, we will use a Bernoulli energy arrival process) and we assume independent arrivals among nodes. However, the model can be easily extended to the more general, time correlated case (e.g., via an underlying common Markov model as in~\cite{Michelusi2013}).

Every node is equipped with a rechargeable battery, so that the energy stored in slot $k$ can be used in a later slot. The global energy level vector in slot $k$ is $\mathbf{e}_k = \langle e_k^1,\ldots,e_k^N \rangle$, and is perfectly known by AP at $k = 0$. This information is used for initializing the parameters of the whole network. After the initial coordination phase at $k = 0$, every node acts independently of the others, and is not aware of the other battery levels in the network.


\subsection{Decentralized--MDPs for EH Systems}\label{subsec:Dec_MDP}

The model presented so far can be formalized using a decentralized Markov decision process framework~\cite{Dibangoye2013}. In our context, an $N$-users Dec-MDP $\mathcal{M} = (\mathbf{E},\mathbf{A},p,r,\eta_0,\beta)$ is formally defined as follows~
\begin{itemize}[leftmargin=*]
    \item \textbf{\emph{Battery Level}} $\mathbf{E} = E^1 \times \cdots \times E^N$ is the set of global battery levels $\mathbf{e} = \langle e^1,\ldots,e^N\rangle$, with $e^i \in E^i \triangleq \{0,\ldots,e_{\rm max}^i\}$ (device $i$ can store up to $e_{\rm max}^i$ discrete energy quanta according to Equation~\eqref{eq:battery_evol}). Throughout, the terms ``battery level'' or ``state'' will be used interchangeably;
    \item \textbf{\emph{Action}} $\mathbf{A} = A^1 \times \cdots \times A^N$ is the set of global actions $\mathbf{a} = \langle a^1,\ldots,a^N\rangle$, where $a^i \in A^i \triangleq [0,1]$ denotes node~$i$'s transmission probability. Although $a^i$ should assume continuous values, we quantize the interval $[0,1]$ in $a_{\rm levels}$ uniformly distributed levels for numerical tractability. Action $a^i$ is chosen by user $i$ through a function $\sigma^i: E^i \to A^i$, and depends only on the local state $e^i$;
    \item \textbf{\emph{Transition Probability}} $p$ is the transition probability function $p: \mathbf{E} \times \mathbf{A} \times \mathbf{E} \to [0,1]$ which defines the probability $p(\bar{\mathbf{e}}|\mathbf{e},\mathbf{a})$ of moving from a global battery level $\mathbf{e} = \langle e^1,\ldots,e^N \rangle$ to a global battery level $\bar{\mathbf{e}} = \langle \bar{e}^1,\ldots,\bar{e}^N \rangle$ under the global action $\mathbf{a}$. When a transmission is performed, one energy quantum is consumed;
    \item \textbf{\emph{Reward}} $r$ is the reward function $r: \mathbf{E} \times \mathbf{A} \to \mathbb{R}^+$ that maps the global action $\mathbf{a}$ to the reward $r(\mathbf{e},\mathbf{a})$ when the global state is $\mathbf{e}$;
    \item $\eta_0$ is the initial state distribution. In our scenario we take~
    \begin{align}\label{eq:eta_0}
        \eta_0(\mathbf{e}) = \begin{cases}
            1, \quad & \mbox{if } \mathbf{e} = \mathbf{e}_0, \\
            0, \quad & \mbox{otherwise},
        \end{cases}
    \end{align}
    
    \noindent for some $\mathbf{e}_0$, i.e., we assume perfect knowledge in the initialization phase.    
    
    \item $1-\beta$ is the probability that the system stops operating in a given slot (see~\cite{Blasco2013}), and will be used in Section~\ref{sec:opt_problem}.
\end{itemize}

In Section~\ref{sec:opt_problem} we describe the optimization problem related to $\mathcal{M}$. Its solution provides a \emph{decentralized control policy}, which will be discussed in Section~\ref{sec:opt_solution}.

Before presenting in more detail the previous bullet points, it is important to emphasize the following key characteristics of the Dec-MDP under investigation:
\begin{itemize}[leftmargin=*]
    \item $\mathcal{M}$ is \emph{jointly fully observable}, i.e., if all nodes collaborated, the global state would be completely known (actually, this is what differentiates Dec-MDPs and Dec-POMDPs);
    \item $\mathcal{M}$ is a \emph{transition independent} Dec-MDP, i.e., the action taken by node $i$ influences only its own battery evolution in that slot and \emph{not} the others. Formally, the transition probability function $p$ can be decomposed as $p(\bar{\mathbf{e}}|\mathbf{e},\mathbf{a}) = \prod_{i = 1}^N p^i(\bar{e}^i|e^i,a^i)$.
    This feature is important to develop compact representations of the transmission policies, and in particular to derive Markovian policies as discussed in our Section~\ref{subsec:reward} and in~\cite[Theorem~1]{Dibangoye2012}.
\end{itemize}

\subsection{Battery Level}

We adopt a discrete model, so that every battery can be referred to as an energy queue, in which arrivals coincide with the energy harvesting process, and departures with packet transmissions. In particular, the battery level of node $i$ in slot $k$ is $e_k^i$ and evolves as~
\begin{align}\label{eq:battery_evol}
    e_{k+1}^i = \min\{e_{\rm max}^i, e_k^i - s_k^i + b_k^i\},
\end{align}

\noindent where the $\min$ accounts for the finite battery size, $s_k^i$ is the energy used for transmission and $b_k^i$ is the energy arrived in slot $k$. $s_k^i$ is equal to $0$ with probability $1-a_k^i$, and to $1$ with probability $a_k^i$.
This model has been widely used in the EH literature~\cite{Michelusi2015}, and represents a good approximation of a real battery when $e_{\rm max}^i$ is sufficiently high. 

\subsection{Action}

Node $i$ can decide to access the channel, with probability $a_k^i$, or to remain idle w.p. $1-a_k^i$. When a transmission is performed, one energy quantum is drained from the battery, and a corresponding reward $g(a_k^i)$ is obtained. When $e_k^i = 0$, no transmission can be performed and $a_k^i = 0$.

\subsection{Transition Probability}

The transition probability function of user $i$, namely $p^i$, is defined as follows (assume $\bar{e} \neq e_{\rm max}^i$)~
\begin{align}
    p^i(\bar{e}|e,a) = \begin{cases}
    (1-p_B^i) a, \quad & \mbox{if } \bar{e} = e-1, \\
    (1-p_B^i) (1-a) + p_B^i a, \quad & \mbox{if } \bar{e} = e, \\
    p_B^i (1-a), \quad & \mbox{if } \bar{e} = e+1, \\
    0, \quad & \mbox{otherwise}.
    \end{cases}
\end{align}

\noindent $p_B^i$ is the probability that user $i$ harvests one energy quantum. More sophisticated models, in which multiple energy quanta can be simultaneously extracted, are described in~\cite{Biason2015d}, and can be integrated in our model (involving, however, higher computational costs).

\subsection{Reward} \label{subsec:reward}

We will use the term ``global reward'' to indicate the overall performance of the system, and simply ``single-user reward'' to refer to the performance of an individual user. We first describe the single-user reward and then extend this to the overall network.

\textbf{\emph{Single-User Reward.}} Assume to study isolated users, which do not suffer from interference, as in~\cite{Michelusi2012}. Data messages are associated with a \emph{potential reward}, described by a random variable $V^i$ which evolves independently over time and among nodes. The realization $\nu_k^i$ is perfectly known only at a time $t \geq kT$ and only to node $i$ whereas, for $t < kT$, only statistical knowledge is available. 
Every node can decide to transmit (and accrue the potential reward $\nu_k^i$) or not in the current slot $k$ according to its value $\nu_k^i$. In particular, it can be shown that a threshold transmission model is optimal for this system~\cite{Michelusi2012}; thus, node $i$ always transmits when $\nu_k^i \geq \nu_{\rm th}^i(e^i)$ and does not otherwise. Note that $\nu_{\rm th}^i(e^i)$ depends on the underlying state (battery level) of user $i$.

On average, the reward of user $i$ in a single slot when the battery level is $e^i$ will be~
\begin{align}
    g(\nu_{\rm th}^i(e^i)) \!\triangleq\! \mathbb{E}[\chi(V^i \!\geq\! \nu_{\rm th}^i(e^i)) V^i] \!=\!\! \int_{\nu_{\rm th}^i(e^i)}^\infty \!\!v f_V^i(v) \ \mbox{d}v,
\end{align}

\noindent where $\chi(\cdot)$ is the indicator function and $f_V^i(\cdot)$ is the pdf of the potential reward, $V^i$.
It is now clear that the transmission probability $a^i$ is inherently dependent on the battery level as~
\begin{align}
    a^i = \sigma^i(e^i) = \int_{\nu_{\rm th}^i(e^i)}^\infty f_V^i(v) \ \mbox{d}v = \bar{F}_V^i(\nu_{\rm th}^i(e^i)),
\end{align}

\noindent where we introduce a function $\sigma^i(e^i)$, which maps local observations ($e^i$) to local actions $\sigma^i(e^i) = a^i$. Note that the complementary cumulative distribution function $\bar{F}_V^i(\cdot)$ is strictly decreasing and thus can be inverted. Therefore, there exists a one-to-one mapping between the threshold values and the transmission probabilities. In the following, we will always deal with $a^i$, and write $g(a^i)$ with a slight abuse of notation.

It can be proved that $g(a^i)$ is increasing and concave in $a^i$, i.e., transmitting more often leads to higher rewards, but with diminishing returns. Finally, note that this model is quite general and, depending on the meaning of $V^i$, can be adapted to different scenarios. For example, in a standard communication system in which the goal is the throughput maximization, $V^i$ can be defined as the transmission rate subject to fading fluctuations~\cite{Michelusi2012}.

\textbf{\emph{Global Reward.}}
The global reward is zero when multiple nodes transmit simultaneously, whereas it is equal to $w^i \nu_k^i$ if only node $i$ transmits in slot $k$ ($w^i$ is the weight of node $i$). On average, since the potential rewards are independent among nodes, we have~
\begin{align}
    r(\nu_{\rm th}(\mathbf{e})) \!=\! \mathbb{E}\bigg[\!\sum_{i = 1}^N w^i V^i \chi(V^i\! \geq\! \nu_{\rm th}^i(e^i)) \prod_{j\neq i} \chi(V^j\! <\! \nu_{\rm th}^j(e^j)) \bigg],
\end{align}
\noindent which can be rewritten as~
\begin{align}\label{eq:r_a}
    r(\mathbf{a}) = r(\boldsymbol{\sigma}(\mathbf{e})) = \sum_{i = 1}^N w^i g(a^i) \prod_{j \neq i} (1-a^j),
\end{align}

\noindent where we used $\mathbf{a}$ instead of $\nu_{\rm th}(\mathbf{e})$ for ease of notation, and we introduced the vector function $\boldsymbol{\sigma} \triangleq \langle \sigma^1,\ldots,\sigma^N \rangle$.  We remark that $\boldsymbol{\sigma}$ summarizes the actions of all users given every battery level, i.e., it specifies all the following quantities~
\begin{align}\label{eq:sigma_struct}
    \begin{matrix}
        \sigma^1(0), & \ldots & \sigma^1(e_{\rm max}^1),\\
        \vdots \\
        \sigma^N(0), & \ldots & \sigma^N(e_{\rm max}^N).
    \end{matrix}
\end{align}

\noindent As we will discuss later, finding $\boldsymbol{\sigma}$ represents the biggest challenge when solving a Dec-MDP.

\begin{figure}[!t]
  \centering
  \includegraphics[trim = 0mm 0mm 0mm 0mm, width=1\columnwidth]{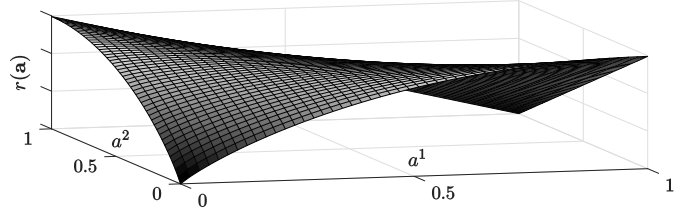}
  \caption{Global reward $r(\mathbf{a})$ when $N = 2$.}
  \label{fig:r_a}
\end{figure}

An important observation is that the reward~\eqref{eq:r_a} is not necessarily increasing nor convex in $\mathbf{a}$, which significantly complicates the solution. An example of $r(\mathbf{a})$ for the two user case can be seen in \figurename~\ref{fig:r_a}. Note that the maximum is achieved when only one device transmits with probability $1$ and the other does not transmit. This implies that, when the devices are not energy constrained (i.e., they have enough energy for transmitting and the current transmission policy does not influence the future), the optimal user allocation should follow an orthogonal approach so as to avoid collisions (the corner points $\langle a^1,a^2 \rangle = \langle 1,0 \rangle$ and $\langle a^1,a^2 \rangle = \langle 0,1 \rangle$ achieve the maximum reward). However, as we will discuss later, this observation does not hold in EH scenarios, in which an action in the current slot influences the future energy levels and, consequently, the future rewards.

Note that, in the previous expressions, we have implicitly restricted our study to Markovian policies. A Markovian policy is a history-independent policy that maps local observations to local actions (i.e., $\sigma^i(e^i) = a^i$). In general decentralized frameworks, tracking previous observations can be used to optimally decide the current action. However, it can be proved~\cite{Dibangoye2012} that under transition independent conditions (which hold in our case, see Section~\ref{subsec:Dec_MDP}), Markovian policies are optimal and thus keeping track of previous states is not necessary.

\section{Optimization Problem} \label{sec:opt_problem}

Ideally, the final goal of the system is to maximize the cumulative weighted discounted long-term reward, defined as~
\begin{align}\label{eq:R_beta_cent}
    \bar{R}_\beta (\pi,\mathbf{e}_0) = \mathbb{E}\left[ \sum_{k = 0}^\infty \beta^k r\Big(\boldsymbol{\sigma}_k(\mathbf{e}_k)\Big) \bigg| \mathbf{e}_0, \pi \right],
\end{align}

\noindent where $\pi \triangleq (\boldsymbol{\sigma}_0,\boldsymbol{\sigma}_1,\ldots)$ is the \emph{policy} and $\beta$ was introduced in Section~\ref{subsec:Dec_MDP} as the probability that the network does not die in a given slot, which corresponds to the \emph{discount factor} in classic MDPs~\cite{Bertsekas2005}. Finding $\pi^\star = \argmax{\pi}\bar{R}_\beta (\pi,\mathbf{e}_0)$ corresponds to obtaining the highest reward when the state of the system $\mathbf{e}_k$ is globally known in slot $k$, i.e., in a \emph{centralized}-oriented network. However, $\bar{R}_\beta(\pi,\mathbf{e}_0)$ cannot be achieved in a decentralized system, thus we must resort to a different notion of long term, which will be given in Equation~\eqref{eq:R_beta_dec}.
Nevertheless, \eqref{eq:R_beta_cent} will be useful to initialize the Dec-MDP solver, since it provides an upper bound to the achievable performance, and can be easily computed using the Value Iteration Algorithm~\cite{Bertsekas2005}.

To formulate the decentralized optimization problem, we first introduce the concept of occupancy state.

\subsection{Occupancy State}

The occupancy state $\eta_k$ is defined as~
\begin{align}\label{eq:eta_k_def}
    \eta_k(\bar{\mathbf{e}}) \triangleq \mathbb{P}(\mathbf{e}_k = \bar{\mathbf{e}} | \eta_0,\boldsymbol{\sigma}_0,\ldots,\boldsymbol{\sigma}_{k-1}),
\end{align}

\noindent and represents a probability distribution over the battery levels given the initial distribution $\eta_0$ and all decentralized decision rules prior to $k$.

It can be shown that the occupancy state represents a sufficient statistic for control purposes in Dec-MDPs, and it can be easily updated at every slot of the system using old occupancy states:~
\begin{align}\label{eq:eta_k_update}
    \eta_k(\bar{\mathbf{e}}) \!=\! \omega(\eta_{k-1},\boldsymbol{\sigma}_{k-1}) \!\triangleq\! \sum_{\mathbf{e}} p(\bar{\mathbf{e}} | \mathbf{e}, \boldsymbol{\sigma}_{k-1}(\mathbf{e})) \eta_{k-1}(\mathbf{e}),
\end{align}

\noindent where $\omega$ is the occupancy update function.
Similarly to the reduction techniques of POMDPs, in which the belief is used as the state in an equivalent MDP, for Dec-MDPs the occupancy state will represent the building block of an equivalent MDP which can be solved using standard techniques.

\subsection{Occupancy-MDP}

Dibangoye \emph{et al.}~\cite{Dibangoye2014} developed a technique to solve Dec-MDPs by recasting them into equivalent continuous state MDPs. In particular, the state space of the equivalent MDP (called \emph{occupancy-MDP}) is the occupancy simplex, the transition rule is given by~\eqref{eq:eta_k_update}, the action space is $\mathbf{A}$, and the instantaneous reward for taking decentralized decision rule $\boldsymbol{\sigma}_k$ is~
\begin{align}\label{eq:rho_def}
    \rho(\eta_k,\boldsymbol{\sigma}_k) = \sum_{\mathbf{e} \in \mathbf{E}} \eta_k(\mathbf{e}) r(\boldsymbol{\sigma}_k(\mathbf{e})),
\end{align}

\noindent i.e., it is the weighted sum of the rewards obtained in every battery level, where the weight is given by the occupancy state. Accordingly, the long-term reward of the occupancy-MDP is~
\begin{align}\label{eq:R_beta_dec}
    R_\beta(\pi,\eta_0) = \mathbb{E}\left[ \sum_{k = 0}^\infty \beta^k \rho(\eta_k,\boldsymbol{\sigma}_k) \bigg| \eta_0,\pi \right].
\end{align}

\noindent Differently from $\bar{R}_\beta(\pi,\eta_0)$ in Equation~\eqref{eq:R_beta_cent}, $R_\beta(\pi,\eta_0)$ can be actually achieved by a decentralized system.

The corresponding optimal policy is~
\begin{align}
    \mu^\star = \argmax{\pi} R_\beta(\pi,\eta_0).
\end{align}

In the next section we discuss how to find the optimal as well as suboptimal solutions.

\section{Solution}\label{sec:opt_solution}

Finding $\mu^\star$ requires solving an MDP with a continuous state space. To do that, we use techniques originally developed for POMDPs which were later used for Dec-POMDPs. In particular, the Learning Real Time A$^*$ (LRTA$^*$) algorithm is suitable for our case, since it explores only the occupancy states which are actually visited during the planning horizon and avoids grid-based approaches (e.g., as used in~\cite{Lovejoy1991}). In~\cite{Dibangoye2012}, the Markov Policy Search (MPS) algorithm was introduced as an adaptation of LRTA$^*$ to decentralized scenarios. In summary, MPS starts at $k = 0$ and uses $\eta_0$ as defined in~\eqref{eq:eta_0}; then, it iteratively updates upper and lower bounds to the optimal policy until they converge to the same value by using the convexity of the cost-to-go function. The solution of the fully-observable MDP of Equation~\eqref{eq:R_beta_cent} is used to initialize the upper bounds at the corner points of the simplex.
For more details about MPS, we refer the readers to~\cite{Dibangoye2012,Dibangoye2013}.

A key step of MPS is the exhaustive backup, in which a new policy that maximizes the cost-to-go upper bound function $\bar{v}_{\beta,k}(\eta_k)$ is obtained. Formally, this requires to solve~
\begin{align}
    \bar{v}_{\beta,k}(\eta_k) = \max_{\boldsymbol{\sigma}} \rho(\eta_k,\boldsymbol{\sigma}) + \beta \bar{v}_{\beta,k+1} (\omega(\eta_k,\boldsymbol{\sigma})).
\end{align}

\noindent In general, the exhaustive backup is critical to perform because all decentralized policies should be examined, thus the complexity would be $\mathcal{O}((a_{\rm levels})^{e_{\rm max}\times N})$ if all users had the same battery size $e_{\rm max}$ (see the structure of $\boldsymbol{\sigma}(\mathbf{e})$ in Equation~\eqref{eq:sigma_struct}). This operation is computationally infeasible when lots of possibilities are involved, thus other approaches were introduced to handle this problem. We first present some preliminary results.

\subsection{Convexity of the Cost-to-go Function}

It can be shown that, in the infinite horizon case, the optimal cost-to-go function $v_{\beta,k}^\star$ is a convex function of the occupancy states (see~\cite[Theorem~4.2]{Dibangoye2013} for a proof in the finite horizon case) and can be approximated by piecewise linear functions. Formally, $\bar{v}_{\beta,k}$ can be rewritten as~
\begin{align}    
    \bar{v}_{\beta,k}(\eta_k)& = \max_{\boldsymbol{\sigma}} \rho(\eta_k,\boldsymbol{\sigma}) + \beta \ C(\Upsilon_k,\omega(\eta_k,\boldsymbol{\sigma})), \label{eq:v_beta_C}
\end{align}

\noindent where $C$ interpolates the occupancy state $\omega(\eta_k,\sigma)$ using the point set $\Upsilon_k$, which contains the visited occupancy states along with their upper bound values. The first points to be put in $\Upsilon_k$ are the corners of the occupancy simplex with their values obtained solving the full knowledge MDP in Equation~\eqref{eq:R_beta_cent}. Then, every time \eqref{eq:v_beta_C} is solved, a new point $(\eta_k,\bar{v}_{\beta,k}(\eta_k))$ is added to $\Upsilon_k$.

Ideally, we could use a linear interpolation as the function $C$ (i.e., map $\eta_k$ on the convex hull of point set $\Upsilon_k$), but this would incur high complexity. A faster solution, which however has shown good performance in many applications, is to replace $C$ with the sawtooth projection:\footnote{The term ``sawtooth'' comes from the shape of the interpolating function in the two-dimensional case. The idea of the approach is to interpolate a point $\eta$ using $|\mathbf{E}|-1$ corner points of the simplex, and one point (the $\ell$-th point) taken from $\Upsilon_k$.}~
\begin{align}\label{eq:sawtooth_def}
    \textbf{sawtooth}(\Upsilon_k&,\!\eta)\! = \! y^0(\eta) \!+\! \min_{\ell \in \mathcal{L}} \!\Big[(v^\ell \!-\!y^0(\eta^\ell)) \min_{\theta:\eta^\ell(\theta)>0} \frac{\eta(\theta)}{\eta^\ell(\theta)}\Big] \nonumber\\
     & =\! y^0(\eta) \!+\! \min_{\ell \in \mathcal{L}} \max_{\theta:\eta^\ell(\theta)>0} \!\Big[\frac{\eta(\theta)}{\eta^\ell(\theta)} (v^\ell \!-\! y^0(\eta^\ell)) \Big] \nonumber \\
     & =\! \min_{\ell \in \mathcal{L}} \bigg[y^0(\eta) \!+\! \max_{\theta:\eta^\ell(\theta)>0} \!\Big[\frac{\eta(\theta)}{\eta^\ell(\theta)} (v^\ell \!-\! y^0(\eta^\ell)) \Big]\bigg].
\end{align}

\noindent $(\eta^\ell,v^\ell)$ is the $\ell$-th element of $\Upsilon_k$, $\mathcal{L}$ is the set of indices of $\Upsilon_k$, and $y^0$ is the upper bound computed using the corner points of $\Upsilon_k$, i.e.,
\begin{align}\label{eq:y_0_def}
    y^0(\eta) = \sum_{\mathbf{e} \in \mathbf{E}} \eta(\mathbf{e}) \Upsilon_k(\mathbf{e}),
\end{align}

\noindent where, with a slight abuse of notation, $\Upsilon_k(\mathbf{e})$ indicates the upper bound value at the corner $\mathbf{e}$ of the simplex.

The sawtooth projection produces higher (i.e., worse) upper bounds than the convex hull projection and thus  MPS may require more iterations to converge (however, a single iteration can be performed much more quickly), but convergence is still guaranteed.

\subsection{Parametric Policies}\label{subsec:sub_opt_parametric}

Since the main issue of the exhaustive backup is that the space of variables is exceedingly large, we aim at reducing this space, so that $\boldsymbol{\sigma}$ cannot take all possible values but is constrained to lie in a smaller subset. This will in turn lead to suboptimal solutions, which however are much faster to compute.
In this paper, we use \emph{parametric policies} and thus reduce the number of optimization variables to few parameters.
In particular, we force the actions of user $i$ to follow a predetermined structure:~
\begin{align}
    \sigma^i(e^i) = f_{\rm par}^i(\Theta^i,e^i)
\end{align}

\noindent where $e^i$ is the independent variable and $\Theta^i$ is a set of parameters which specify the structure of $f_{\rm par}^i$. For example, if we used $\Theta^i = \{\theta^i\}$, and a simple linear function $f_{\rm par}^i(\Theta^i,e^i) = \theta^i e^i$, the only optimization variable of user $i$ would be $\theta^i$, and not $\sigma^i(0),\ldots,\sigma^i(e_{\rm max}^i)$ as in the original problem.
In this case, for a symmetric scenario, the complexity of the exhaustive search step goes from $\mathcal{O}((a_{\rm levels})^{e_{\rm max} \times N\vphantom{|}})$ to $\mathcal{O}((\theta_{\rm levels})^{N\vphantom{!}})$, therefore it remains exponential in $N$ but with a much smaller coefficient in the exponent. $\theta_{\rm levels}$ is the number of values that $\theta^i$ can assume. 

In our scenario we force $f_{\rm par}^i(\Theta^i,e^i)$ to be a non-decreasing function of $e^i$ as in~\cite{Michelusi2012}, which implies that higher energy levels cannot correspond to lower transmission probabilities.

\section{Numerical Evaluation}\label{sec:num_eval}

\begin{figure}[!t]
  \centering
  \includegraphics[trim = 0mm 0mm 0mm 0mm, width=.95\columnwidth]{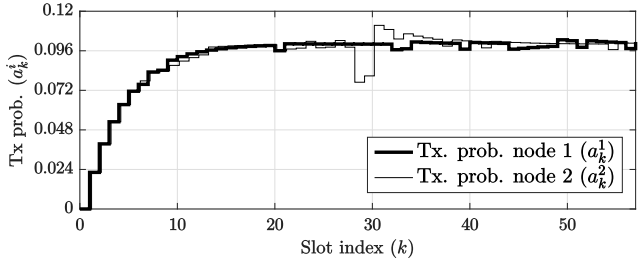}
  \caption{Transmission probabilities as a function of time for two users with batteries initially empty.}
  \label{fig:tx_prob_discharged}
\end{figure}
\begin{figure}[!t]
  \centering
  \includegraphics[trim = 0mm 0mm 0mm 0mm, width=.95\columnwidth]{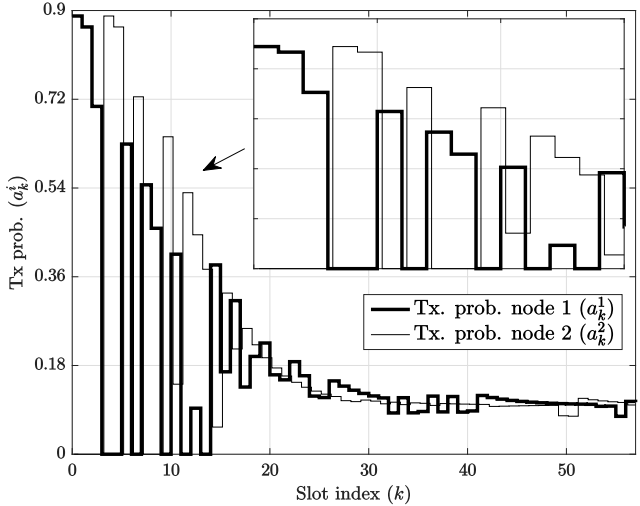}
  \caption{Transmission probabilities as a function of time for two users with batteries initially fully charged.}
  \label{fig:tx_prob_fullyCharged}
\end{figure}

The numerical evaluation is performed using two nodes, since the complexity grows super-exponentially with the number of users. Indeed the size of the occupancy state evolves exponentially with $N$, and the exhaustive backup operation (exponential in $N$), or a suboptimal approach, is to be performed for every element of the occupancy state. If not otherwise stated, we adopt the following parameters: the batteries can contain up to $e_{\rm max}^1 = e_{\rm max}^2 = 5$ energy quanta; the probabilities of receiving an energy quantum are equal to $p_B^1 = p_B^2 = 0.1$ in every slot (i.i.d. energy arrival processes); 
when a transmission is performed a reward $V^i = \ln(1+\Lambda^i H^i)$ is accrued, where $V^i$ is defined as the normalized transmission rate in a slot, where $H^i$ is an exponentially distributed random variable with mean $1$ (see~\cite{Michelusi2012}); the average normalized SNRs are $\Lambda^1 = 6$ and $\Lambda^2 = 4$; both devices have the same weight; finally, the discount factor is $\beta = 0.9$. All the numerical evaluations were written in C++.

In \figurename s~\ref{fig:tx_prob_discharged} and~\ref{fig:tx_prob_fullyCharged} we show the transmission probabilities of the parametric decentralized policy of Section~\ref{subsec:sub_opt_parametric}, where $f_{\rm par}$ is a linear function, $\Theta^i = \{\theta^i\}$ and $\theta^i$ is such that $\theta^i e_{\rm max}^i \in A^i$.

From these two figures, an interesting effect can be observed in the initial transmission slots: when the available energy is scarce, then both nodes have a non-zero probability of accessing the channel; instead, if a lot of energy is available, the transmission policy almost degenerates into a pure time-orthogonal access mechanism. Also, in \figurename~\ref{fig:tx_prob_discharged}, the average transmission probabilities coincide with the energy arrival rate in the long run, so as to achieve energy neutrality. In summary, this proves that when the energy resources are scarce (here we show this effect in terms of initial energy levels, but a similar behavior could be observed for low energy arrivals) then an orthogonal scheme, in which collisions are avoided, is suboptimal. The trade-off between orthogonal and random access schemes can be intuitively explained as follows. When the initial energy levels are high but the harvesting probabilities are low, both nodes know that the other device has a lot of energy available in the first slots. Thus, since there are no energy outages, they can adopt an orthogonal access policy, so that the channel is almost always used (which corresponds to the best mechanism without EH constraints). This regime almost corresponds to the full-knowledge case. Instead, as time goes on, nodes lose information about the global state of the system, thus a device does not know the energy level of the other. In this case, an orthogonal scheme might be highly inefficient, since a node may not have enough energy to transmit during its slots; here, a random access scheme provides higher performance.

\begin{figure}[!t]
  \centering
  \includegraphics[trim = 0mm 0mm 0mm 0mm, width=.95\columnwidth]{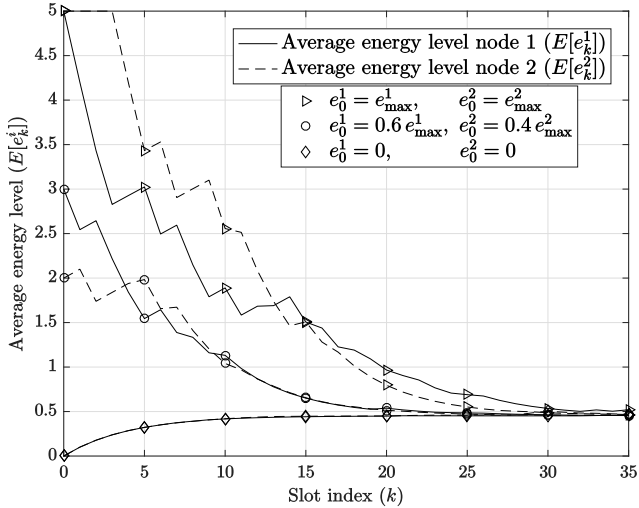}
  \caption{Battery level evolution as a function of time for two users with different initial battery levels.}
  \label{fig:energy_level}
\end{figure}

\figurename~\ref{fig:energy_level} shows another interesting, although predictable, result: regardless of the initial energy level, in the long run all the energy levels degenerate to the same value. This is because all the initial fluctuations have been absorbed by the batteries. Moreover, we also remark that, after many slots, the knowledge of a user about the others coincides with their steady-state probabilities, since the global battery knowledge is not refreshed at any time after $k = 0$.

\begin{figure}[!t]
  \centering
  \includegraphics[trim = 0mm 0mm 0mm 0mm, width=.95\columnwidth]{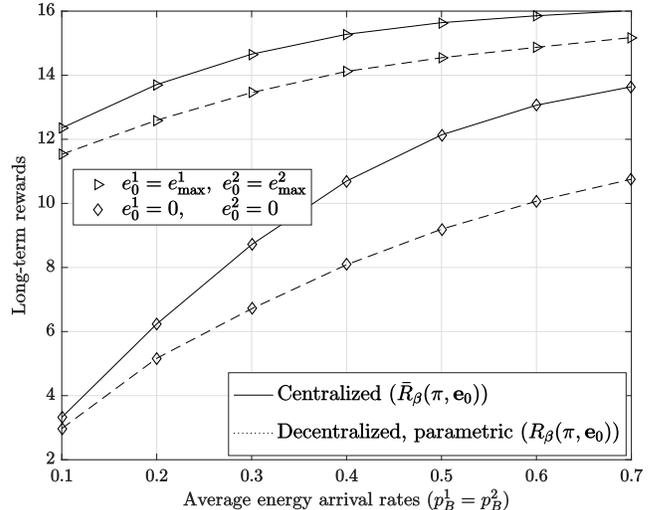}
  \caption{Centralized and decentralized long-term rewards as a function of the energy arrival rates for two users with different initial battery levels.}
  \label{fig:rewards}
\end{figure}

Finally, in \figurename~\ref{fig:rewards} we show the long-term discounted reward as a function of the energy arrival rate for the optimal centralized scheme (Equation~\eqref{eq:R_beta_cent}) and the decentralized parametric scheme (Equation~\eqref{eq:R_beta_dec}). When the initial batteries are fully charged, then centralized and decentralized schemes are much closer, whereas, for battery initially empty, the gap is much wider.

Another counter-intuitive phenomenon can be observed as the average energy arrival rate grows. Indeed, as previously explained, when a lot of energy is available, an almost orthogonal scheme is optimal, thus centralized and decentralized schemes should have similar performance. However, in \figurename~\ref{fig:rewards} the opposite effect can be observed. This is because we are using a discounted formulation, thus the first slots are the most important ones. When a lot of energy arrives to the system, the battery fluctuations are more frequent, thus the distance between centralized and decentralized approaches becomes wider.
Finally, note that the performance of the parametric policy is strongly influenced by the number of parameters $\Theta^i$ we used, and using more parameters would lead to better performance.

\section{Conclusions}\label{sec:conclusions}

We studied a decentralized optimization framework for an energy harvesting communication network with collisions. We used a decentralized Markov decision process to model the system, and described how to find the optimal policy as well as suboptimal schemes. In our numerical evaluation we describe the trade-off between accessing the channel and energy arrivals, and we showed that a pure orthogonal access mechanism is suboptimal under harvesting constraints.

Due to the super-exponential complexity of the optimal solution, future work will investigate more practical schemes which inherit the key properties of our framework while being less computational demanding.

\bibliography{bibliography}{}
\bibliographystyle{IEEEtran}

\end{document}